\documentclass[reprint,aps,ams,amsmath,prx,twocolumn,longbibliography]{revtex4-1}
\usepackage{graphics}
\usepackage{graphicx}
\usepackage{epsfig}
\usepackage{amsmath}
\usepackage{amssymb}
\usepackage{amsfonts}
\usepackage{bm}
\usepackage{color}
\usepackage{bbm}
\usepackage{hyperref}
\usepackage[normalem]{ulem}
\usepackage{comment}
\usepackage{soul}
\usepackage[OT1]{fontenc}

\begin{document}

\title{Nonreciprocal Coulomb drag in electron bilayers}

\author{Dmitry Zverevich}
\affiliation{Department of Physics, University of Wisconsin-Madison, Madison, Wisconsin 53706, USA}

\author{Alex Levchenko}
\affiliation{Department of Physics, University of Wisconsin-Madison, Madison, Wisconsin 53706, USA}

\date{July 23, 2025}

\begin{abstract}
We propose a mechanism and develop a theory for nonreciprocal Coulomb drag resistance. This effect arises in electron double-layer systems in the presence of an in-plane magnetic field in noncentrosymmetric conductors or in bilayers with spontaneously broken time-reversal symmetry and without Galilean invariance. We demonstrate the significance of this effect
by examining the hydrodynamic regime of the electron liquid. The nonreciprocal component of the transresistance is shown to sensitively depend on the intrinsic conductivity, viscosity of the fluid, and the emergent nonreciprocity parameter.   
\end{abstract}

\maketitle

Onsager’s reciprocity relations \cite{Onsager-I,*Onsager-II} and the fluctuation-dissipation theorem established by Callen and Welton \cite{FDT:1951} are cornerstones of linear transport theory. An equivalent statement of the theorem is that any two-terminal conductance in a system can only have an even dependence on the magnetic field. However, the symmetry constraints imposed on linear conductance do not generally extend to the nonlinear regime, particularly in systems with broken symmetries, most commonly time-reversal and inversion. Electrical magnetochiral anisotropy \cite{Rikken:2001} is a notable example of a nonlinear magnetotransport effect that violates Onsager’s principle, where resistance depends linearly on both the external magnetic field and the current through the conductor. In the context of mesoscopic electron transport in disordered and ballistic systems, odd-in-field corrections to the nonlinear current-voltage characteristics are known to be highly sensitive to the strength of electron-electron interactions \cite{Spivak:2004,Sanchez:2004,Tsvelik:2006,Andreev:2006,Deyo:2006}. In this work, we present results on an analogous nonreciprocal effect in the context of interactively coupled transport in electron double-layer systems, driven by interlayer Coulomb interactions.

The effect in question pertains to Coulomb drag, a phenomenon that serves as a sensitive experimental tool for probing electronic coherence and correlations. The microscopic mechanisms underlying drag resistance (or equivalently, transresistance) are well understood and, in general, agree well with experimental measurements; see Ref. \cite{Review:2016} for a review and references therein. In one physical interpretation \cite{Kamenev:1995}, Coulomb drag can be viewed as a second-order nonequilibrium rectification effect, where thermally driven electron density fluctuations in one layer, advected by the flow of the electron liquid in the presence of an electric field, induce a dc response in the other layer through interlayer Coulomb interactions. Existing results establish a connection between drag and the current noise power spectrum density, making a critical step forward in understanding the nonlinear regime and fluctuation relations far from equilibrium \cite{Levchenko:2008,Chudnovskiy:2009,Sanchez:2010,Sukhorukov:2019}. Recent experiments demonstrate observed notable nonreciprocal features of Coulomb drag in coupled 1D quantum wires \cite{Laroche:2024}, carbon nanotube/nanowire-monolayer graphene hybrid devices \cite{Das:2020,Anderson:2021}, graphene moir\'e heterostructures \cite{Ki:2024}, and thin films of topological anomalous Hall bilayers \cite{Fu:2025}. However, no established theory exists for any of these systems, and more broadly, the possible microscopic origins of nonreciprocal Coulomb drag remain unexplored--an open question that, in part, motivates this study. 

\begin{figure}[t!]
\includegraphics[width=\linewidth]{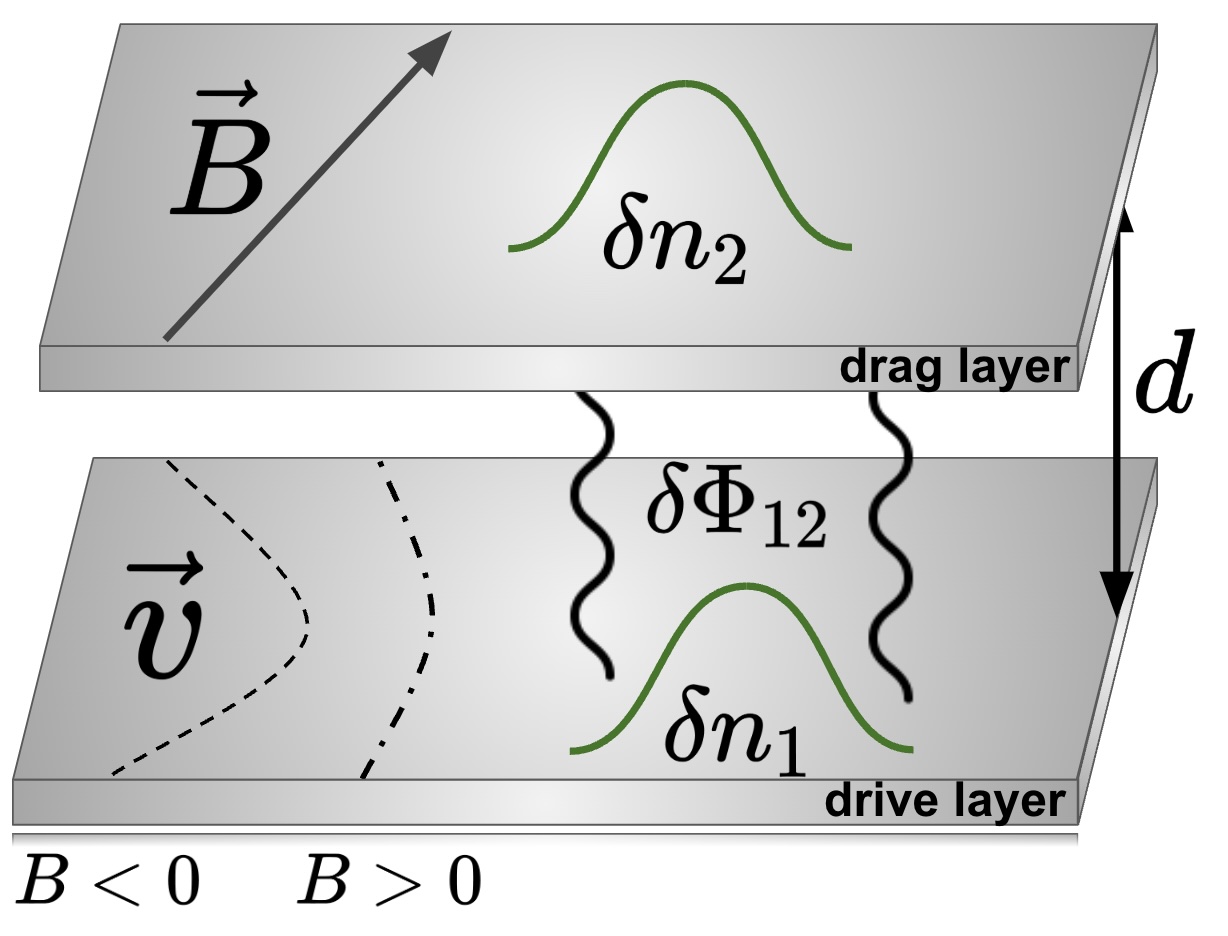}
\caption{A schematic representation of an electron bilayer in a Coulomb drag setup. The nonreciprocal character of electron flow in the drive layer with velocity $\bm{v}$ is illustrated for different in-plane magnetic field orientations. The interlayer Coulomb potential $\delta\Phi_{12}$ couples density fluctuations $\delta n_{1,2}$, resulting in a drag force between the layers.}\label{fig:CD}
\end{figure}

As any other transport coefficient drag resistivity is sensitive to different types of electron scattering. In our considerations of the Coulomb drag setup consisting of two 2D conductors separated by a distance $d$, see Fig. \ref{fig:CD} for illustration, we assume that both layers are in the clean limit, meaning that the intralayer mean free path due to electron-electron collisions, $l_{\text{ee}}$, 
is much shorter than the other relevant scattering length scales -- namely, the electron-impurity mean free path $l_{\text{ei}}$ and the electron-phonon mean free path $l_{\text{ep}}$, which govern the relaxation of electron momentum and energy. In this limit, the electron liquid in each layer can be described within the hydrodynamic approximation. The feasibility of reaching this regime for electron liquids is well established in high-mobility semiconductors and graphene devices, see reviews on the topic in Refs. \cite{Lucas:2018,Levchenko:2020,Narozhny:2022,Fritz:2024}.
Hydrodynamic effects in Coulomb drag resistance were already reported in several experiments \cite{Titov:2013,Dean:2016,Ponomarenko:2024}.   
  
 In one layer (the drive layer) we assume a steady flow of the electron fluid with hydrodynamic velocity $\bm{v}$. The other layer (the drag layer) is subject to open-circuit conditions, meaning no net current flows through it. In the following, we work in the limit $k_{\text{F}}d\gg1$, where $k_{\text{F}}$ is the corresponding Fermi momentum \footnote{We work in the physical units by setting Planck's constant and Boltzmann's constant unity $\hbar=k_{\text{B}}=1$}. For simplicity, we assume that the average electron density $n$ is the same in both layers. We begin our technical discussion by outlining the main equations within the hydrodynamic framework and then solving them to determine the drag resistivity.

The thermally-driven spatial and temporal density fluctuations of the electron density $\delta n_a(\bm{r},t)$ in each layer $a=1,2$ are linked to the corresponding velocity fluctuations of the fluid $\delta\bm{v}_a(\bm{r},t)$ through the continuity equation.  We linearize this equation in small fluctuations $\delta n\ll n$ and $\delta\bm{v}\ll\bm{v}$ and in the drive layer ($a=1$) it takes the form 
\begin{equation}\label{eq:continuity}
\partial_t\delta n_1+\bm{v}\cdot\bm{\nabla}\delta n_1+n\bm{\nabla}\cdot\delta\bm{v}_1-\frac{\sigma}{e^2}\bm{\nabla}^2e\delta\Phi_{12}+\bm{\nabla}\cdot\delta\bm{I}_1=0.
\end{equation}
In the drag layer $a=2$ it has the same form with $\bm{v}\to0$ and flipped indices $1\leftrightarrow2$. The first three terms are conventional coming from the linearization of $\partial_tn$ and $\bm{\nabla}(n\bm{v})$. The third term arises in systems with broken Galilean invariance. It describes current fluctuations due to finite intrinsic conductivity $\sigma$ \cite{Patel:2017}. They are rendered by fluctuations of the Coulomb potential $\delta\Phi_{12}$, which is related by the Poisson equation to density fluctuations in both layers. In the Fourier components it admits a simple form 
\begin{equation}
e\delta\Phi_{ab}(\bm{q},\omega)=\frac{2\pi e^2}{\varkappa q}\left[\delta n_a(\bm{q},\omega)+\delta n_b(\bm{q},\omega) e^{-qd}\right],
\end{equation}
where $\varkappa$ is the dielectric constant of the interlayer material. The last term in Eq. \eqref{eq:continuity}, is the random flux of current (Langevin extraneous noise) whose correlation function is established by the fluctuation-dissipation relation       
\begin{equation}\label{eq:I-I}
\langle \delta I_{ia}(\bm{r},t)\delta I_{jb}(\bm{r}',t')\rangle=2T\frac{\sigma}{e^2}\delta(\bm{r}-\bm{r}')\delta(t-t')\delta_{ij}\delta_{ab}, 
\end{equation}
where indices $i,j$ denote Cartesian coordinates for current components within each layer and the brackets $\langle\ldots\rangle$ denote averaging over the thermal fluctuations.  

The second relation between $\delta n_a$ and $\delta \bm{v}_b$ is established through the momentum continuity equation for the electron liquid in each layer. It  can be expressed in the form of Newton’s second law, which, in turn, takes the form of the Navier-Stokes equation for a charged fluid. In its linearized form, it reads for the drive layer 
\begin{equation}\label{eq:NS}
\varrho(\partial_t+\bm{v}\cdot\bm{\nabla})\delta\bm{v}_1=-\bm{\nabla}\cdot\delta\hat{\Pi}_1-en\bm{\nabla}\delta\Phi_{12},
\end{equation}
where $\varrho$ is the mass density of the fluid. The second term on the right-hand side of this equation describes the flow of momentum in the electron fluid due to long-range Coulomb interactions between electrons. In contrast, the first term on the right-hand side of Eq. \eqref{eq:NS} represents the local contribution to fluctuations in the momentum flux tensor
\begin{equation}
\delta\hat{\Pi}\equiv\delta\Pi_{ij}=\delta P \delta_{ij}-\delta\Sigma_{ij},
\end{equation}
which consists of pressure fluctuations $\delta P$ in the fluid, where $\delta_{ij}$ is Kronecker delta symbol, and fluctuations in viscous stresses
\begin{equation}\label{eq:Sigma}
\delta\Sigma_{ij}=\delta\Sigma^e_{ij}+\delta\Sigma^o_{ij}+\delta\Sigma^n_{ij}+\delta\Xi_{ij}.
\end{equation}
The latter can be split into several parts. The first one (even term) takes the usual form
\begin{equation}
\delta\Sigma^e_{ij}=2\eta\delta V_{ij}+(\eta-\zeta)\delta_{ij}\bm{\nabla}\cdot\delta\bm{v},
\end{equation}
where variation of the strain rate is $\delta V_{ij}=(\partial_i\delta v_j+\partial_j\delta v_j)/2$, with $\eta$ and $\zeta$ denoting shear and bulk viscosities of the fluid. Here 
we used shorthand notation for the spatial derivative $\partial_i=\partial/\partial x_i$. In situations where time-reversal symmetry is broken either spontaneously or by an external magnetic field, the stress tensor acquires additional odd components, whose fluctuations are given by
\begin{equation}
\delta\Sigma^o_{ij}=\eta_o(\epsilon_{ik}\delta V_{jk}+\epsilon_{jk}\delta V_{ik}), 
\end{equation} 
where $\epsilon_{ij}$ is the 2D antisymmetric Levi-Civita tensor, and $\eta_o$ is the dissipationless odd viscosity \cite{Avron:1998}. 
The most crucial term for our problem is the third one that captures the nonreciprocal part of the stress tensor. As discussed recently \cite{Kawakami:2021,Kirkinis:2025}, it receives contributions from various effects including the quantum geometric tensor. The key insight that we apply here is that in systems without time-reversal symmetry and Galilean invariance viscosity may depend on the flow velocity. Therefore, from the strain rate $V_{ij}$, which is a rank-2 tensor, and a vector $\bm{v}$, one can construct the stress tensor 
by introducing a rank-5 nonreciprocity tensor 
\begin{equation}
\delta\Sigma^n_{ij}=\eta N_{ijklm}v_k\delta V_{lm}. 
\end{equation}     
The specific structure of this tensor depends on the nature of symmetry breaking in the system. We consider the simplest scenario, applicable to a noncentrosymmetric conductor under an in-plane magnetic field $\bm{B}$. Several tensor combinations can form $N_{ijklm}$, constructed from $\delta_{ij}$, $\epsilon_{ij}$, and $B_i$. For brevity, we focus on a particular contribution by taking $N_{ijklm}=\alpha\delta_{il}\delta_{jm}B_k$, where $\alpha$ is a phenomenological parameter whose value must be determined from a microscopic theory. This choice does not restrict our conclusions, as the calculation can be generalized to other cases, leading to the same result with a redefined nonreciprocity parameter, which we introduce below. Finally, the last term in Eq. \eqref{eq:Sigma} accounts for stochastic fluxes arising from thermally driven fluctuations of viscous stresses, whose correlation function is determined by the fluctuation-dissipation theorem \cite{LL:1957}
\begin{align}\label{eq:Xi}
\langle\delta\Xi_{ik}(\bm{r},t)\delta\Xi_{lm}(\bm{r}',t')\rangle=2T\delta(\bm{r}-\bm{r}')\delta(t-t')\nonumber \\ 
\times [\eta(\delta_{il}\delta_{km}+\delta_{im}\delta_{kl})+(\eta-\zeta)\delta_{ik}\delta_{lm}].
\end{align}
The odd viscosity does not enter here as it is a nondissipative property of the fluid and we suppressed possible nonreciprocal corrections, which would lead to a higher-order effects.   

In the presence of the steady current $\bm{j}_1=en\bm{v}$ the drive layer exerts a force $\bm{F}_{\text{D}}$ on the drag layer due to Coulomb coupling of density fluctuations between the layers \footnote{Note that at sufficiently high density the part of the current in the drive layer due to the intrinsic conductivity can be neglected in comparison to the convective part of the current}. It can be determined by relating the potential to density fluctuations by using the Poisson equation. In the Fourier components it takes the form \cite{Apostolov:2014}
\begin{equation}\label{eq:F}
\bm{F}_{\text{D}}=\int\frac{d^2qd\omega}{(2\pi)^3}(-i\bm{q})\left(\frac{2\pi e^2}{\varkappa q}\right)e^{-qd}D(\bm{q},\omega),
\end{equation}     
that is expressed through the dynamical structure factor  
\begin{equation}\label{eq:D}
D(\bm{q},\omega)=\langle\delta n_1(\bm{q},\omega)\delta n_2(-\bm{q},-\omega)\rangle,
\end{equation}
which is the interlayer correlation function of density fluctuations. In the open circuit of the drag layer this force is balanced by the built-in electric field $\bm{E}_2$, therefore the force balance condition applied on an element of the fluid is $\bm{F}_{\text{D}}=en\bm{E}_2$. The drag resistivity $\rho_{\text{D}}$ is defined as the ratio of this generated field induced by the current in the other layer $\bm{E}_2=\rho_{\text{D}}\bm{j}_1$. As a result, it can be expressed through the drag force 
\begin{equation}\label{eq:rho}
\bm{F}_{\text{D}}=(en)^2\rho_{\text{D}}\bm{v}.
\end{equation}     

To determine $\bm{F}_{\text{D}}$, we must solve the coupled equations \eqref{eq:continuity} and \eqref{eq:NS} for both layers, resulting in a total of four differential equations. We approach this by applying a Fourier transform to all fluctuating quantities, $\delta n_a,\delta\bm{v}_a,\delta\Phi_{ab}\propto e^{-i\omega t+i\bm{qr}}$.
Using the continuity equations \eqref{eq:continuity}, we express the velocity fluctuations $\delta\bm{v}_a$ in terms of the density fluctuations $\delta n_a$. Substituting these expressions into the Navier-Stokes equations \eqref{eq:NS} yields a closed set of equations for the density fluctuations alone. While these equations can be solved analytically, the general solution is excessively cumbersome. Instead, we simplify the problem by analyzing the relevant energy scales and their roles.
We observe that hydrodynamic electron density fluctuations propagate as plasmons \cite{Zverevich:2023}. In a bilayer system, there are two branches of plasma oscillations: the in-phase optical plasmon and the out-of-phase acoustic plasmon. Their dispersion relations are given by
\begin{equation}
\omega_\pm(q)=\omega_{\text{p}}\sqrt{1\pm e^{-qd}},\quad \omega_p=\sqrt{\frac{2\pi n^2e^2q}{\varrho\varkappa}}.
\end{equation}        
Next we observed that there are two energy scales due to viscous effects in the fluid which are given by
\begin{equation}
\omega_\nu=\nu q^2,\quad \omega_o=\nu_o q^2,
\end{equation}
where $\nu=(\eta+\zeta)/\varrho$ and $\nu_o=\eta_o/\varrho$ are usual and odd kinematic viscosities. At charge neutrality, the $\omega_o$ mode corresponds to the odd viscosity wave. At finite electron density it hybridizes with plasmons to give a collective mode with the dispersion $\sqrt{\omega^2_\pm+\omega^2_o}$. Since plasmon modes scale with momentum as $\omega_+\propto\sqrt{q}$ and $\omega_-\propto q$ the correction $\propto q^2$ due to odd viscosity can be omitted for density fluctuations with long wavelength. The conventional viscous term $\omega_\nu$ makes plasmon energies complex, meaning it gives them finite life time, so that density oscillations decay in time $\propto e^{-\nu q^2t}$. In addition to that, due to finite intrinsic conductivity there is also and energy scale 
\begin{equation}
\gamma_\pm(q)=\gamma_q(1\pm e^{-qd}),\quad \gamma_q=\frac{2\pi\sigma q}{\varkappa},
\end{equation}       
which also contributes to the attenuation of plasmons. Physically it corresponds to the Maxwell mechanism of charge relaxation. 
It is clear that in the long-wavelength limit, $qd\ll1$, plasmon damping is dominated by Maxwell relaxation due to the intrinsic conductivity, since $\gamma_\pm\gg\omega_\nu$. 

These considerations suggest the following sensible approximation: in the Navier-Stokes equations, viscosity can be set to zero everywhere except in the nonreciprocal term, as plasmon decay is primarily governed by the Maxwell mechanism, while nonreciprocity arises from flow-dependent viscosity. Additionally, in the long-wavelength limit, pressure gradients in the momentum flux tensor can be neglected, as they are subleading compared to the Coulomb term.
This leads us to the compact form of equations determining density fluctuations
\begin{align}\label{eq:n-n}
\mathcal{P}_\pm\delta n_\pm=-i\omega^2(\bm{q}\cdot\delta\bm{I}_\pm)+\frac{i}{2}(\bm{q}\cdot\bm{v})[\Gamma_+\delta n_++\Gamma_-\delta n_-]\nonumber \\ 
-2\alpha\omega_\nu(\bm{v}\cdot\bm{B})[\omega^2_+\delta n_++\omega^2_-\delta n_-]
\end{align}
where
$\mathcal{P}_\pm(q,\omega)=i\omega\Gamma_\pm(q,\omega)+\omega^2\gamma_\pm$,  
$\Gamma_\pm(q,\omega)=\omega^2_\pm-\omega^2$,
and we introduced symmetrized notations $\delta n_\pm=\delta n_1\pm\delta n_2$ and $\delta\bm{I}_\pm=\delta\bm{I}_1\pm\delta\bm{I}_2$. Poles of $\delta n$, zeros of $\mathcal{P}_\pm$, define dispersion laws for collective modes in the systems. As discussed above these are plasmons. $\Gamma_\pm$ can be interpreted as vertex functions that couple nonequilibrium density fluctuations. Equations \eqref{eq:n-n} can be solved perturbatively in $\bm{v}$ and we present the result in a form of a series 
\begin{subequations}
\begin{align}
&\delta n_\pm=\delta n^{(0)}_\pm+\delta n^{(1)}_\pm+\delta n^{(2)}_\pm, \\
&\delta n^{(0)}_\pm=\frac{-i\omega^2}{\mathcal{P}_\pm}(\bm{q}\cdot\delta\bm{I}_\pm),\\
&\delta n^{(1)}_\pm=\frac{i}{2\mathcal{P}_\pm}(\bm{q}\cdot\bm{v})\left[\Gamma_+\delta n^{(0)}_++\Gamma_-\delta n^{(0)}_-\right],\\
&\delta n^{(2)}_\pm=-\frac{2\alpha\omega_\nu}{\mathcal{P}_\pm}(\bm{v}\cdot\bm{B})\left[\omega^2_+\delta n^{(1)}_++\omega^2_-\delta n^{(1)}_-\right].
\end{align}
\end{subequations}
In this form each term has a transparent physical meaning. The first term in the series represents equilibrium density fluctuations in the system. The second term accounts for the nonequilibrium component advected by the flow. The third term corresponds to the nonreciprocal contribution to the nonequilibrium fluctuations.

\begin{figure}[t!]
\includegraphics[width=\linewidth]{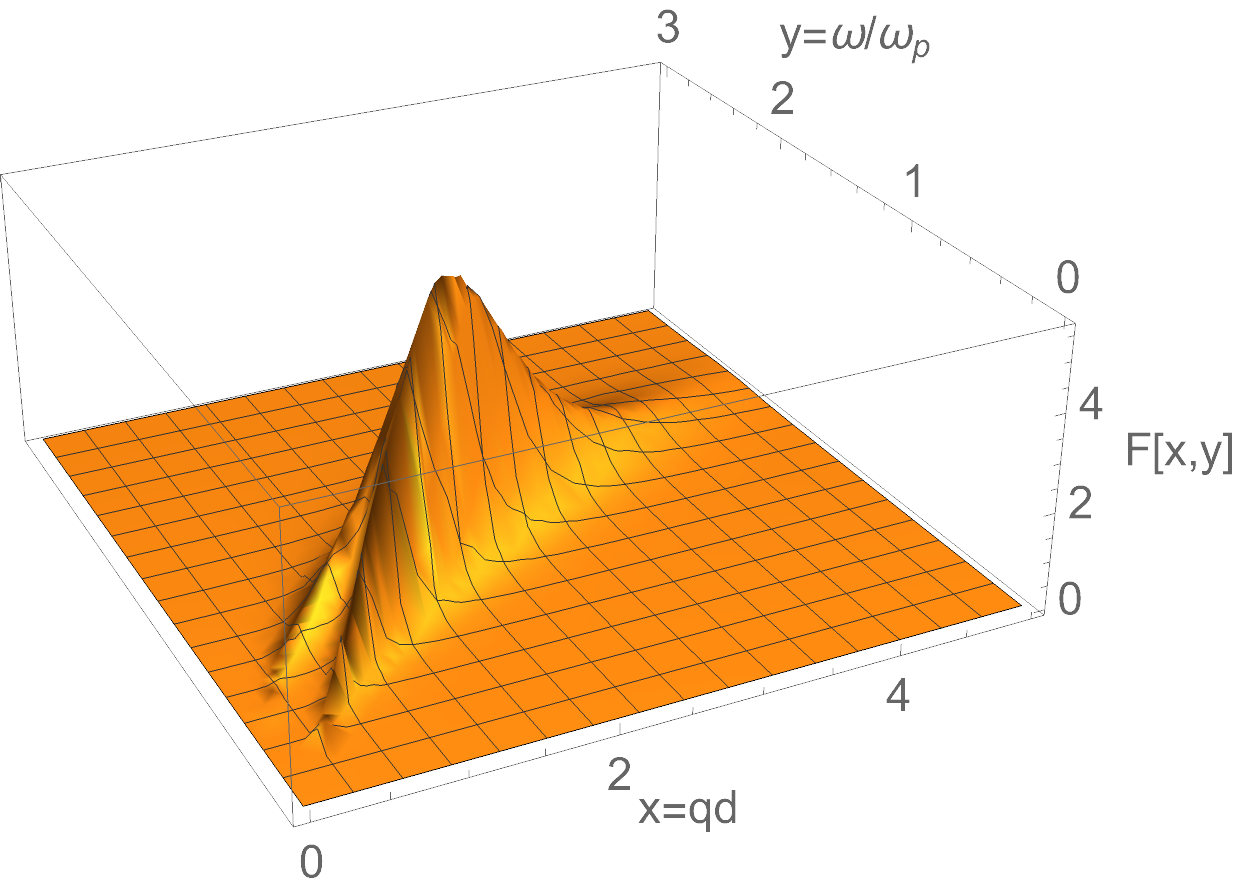}
\caption{The spectral density plot of the integrand of Eq. \eqref{eq:F} in the rescaled units $x=qd$ and $y=\omega/\omega_p$ representing the spectral weight of a function that defines linear part of the drag force $\bm{F}_{\text{D}}$. It depends on a single parameter $\varsigma=\gamma_q/\omega_p$ taken at $q=1/d$. On the plot we took $\varsigma=0.15$. This pot is qualitatively similar for all values of this parameter in the range $\varsigma<1$. }\label{fig:3D}
\end{figure}

We proceed to calculate the dynamical structure factor defined by Eq. \eqref{eq:D}. For this purpose we use the correlation function in Eq. \eqref{eq:I-I}, which in the Fourier components of the symmetrized basis reads $\langle(\bm{q}\cdot\delta\bm{I}_\pm)(\bm{q}\cdot\delta\bm{I}_\pm)\rangle=2T\sigma q^2/e^2$. The equilibrium part of $D(\bm{q},\omega)$ obviously does not contribute to the drag force, so we simply omit it. Its nonequilibrium parts are found in the form 
\begin{subequations}\label{eq:D-1-2}
\begin{align}
&D(\bm{q},\omega)=D^{(1)}(\bm{q},\omega)+D^{(2)}(\bm{q},\omega),\\
&D^{(1)}(\bm{q},\omega)=i\Upsilon\frac{\Gamma_+\Re\mathcal{P}_--\Gamma_-\Re\mathcal{P}_+}{|\mathcal{P}_+|^2|\mathcal{P}_-|^2},\\
&D^{(2)}(\bm{q},\omega)=4i\chi\Upsilon\omega_\nu\gamma_qe^{-qd}\frac{\omega^2_+\gamma_+|\mathcal{P}_-|^2+\omega^2_-\gamma_-|\mathcal{P}_+|^2}{|\mathcal{P}_+|^4|\mathcal{P}_-|^4}
\end{align}
\end{subequations} 
Here for compactness we introduced the notation for $\Upsilon=(\bm{q}\cdot\bm{v})T\frac{\sigma}{e^2}q^2\omega^4$ and the dimensionless nonreciprocity parameter $\chi\propto (\bm{j}\cdot\bm{B})$ defined precisely below in Eq. \eqref{eq:variables}. 

To correlate our discussion of collective modes and characteristic energy scales with the obtained spectral density of fluctuations, it is useful to express the results in appropriate units. The natural scale for $q$ is set by the inverse interlayer separation, $1/d$, while the natural energy scale is given by the plasma frequency  $\omega_{\text{p}}$ at $q=1/d$. Accordingly, we introduce the dimensionless variables $x=qd$ and $y=\omega/\omega_p$. Figure \ref{fig:3D} displays the integrand of Eq. \eqref{eq:F}, namely, the unintegrated force density $F(x,y)$, as a function of $x$ and $y$, scaled by an overall factor. This integrand is determined by the product of the dynamical structure factor, the screened interlayer Coulomb potential, the strength of Langevin fluxes, and the phase space volume. The plasmon resonances manifest as pronounced peaks along the lines $y=\sqrt{x(1\pm e^{-x})}$. Overall, the force is primarily governed by values around $x\sim y\sim1$. At large $q$ (high $x$), the plasmon branches merge as their energy splitting becomes exponentially small. The broadening of these resonant features is controlled by the ratio of the plasmon attenuation coefficient $\gamma_q$ to the plasma frequency $\omega_p$ at $q=1/d$. At very small $q$ and $\omega$, resonant features are suppressed due to a combination of phase space constraints and the strong $q,\omega$ dependence of the vertex functions $\Gamma_\pm$.

\begin{figure}[t!]
\includegraphics[width=\linewidth]{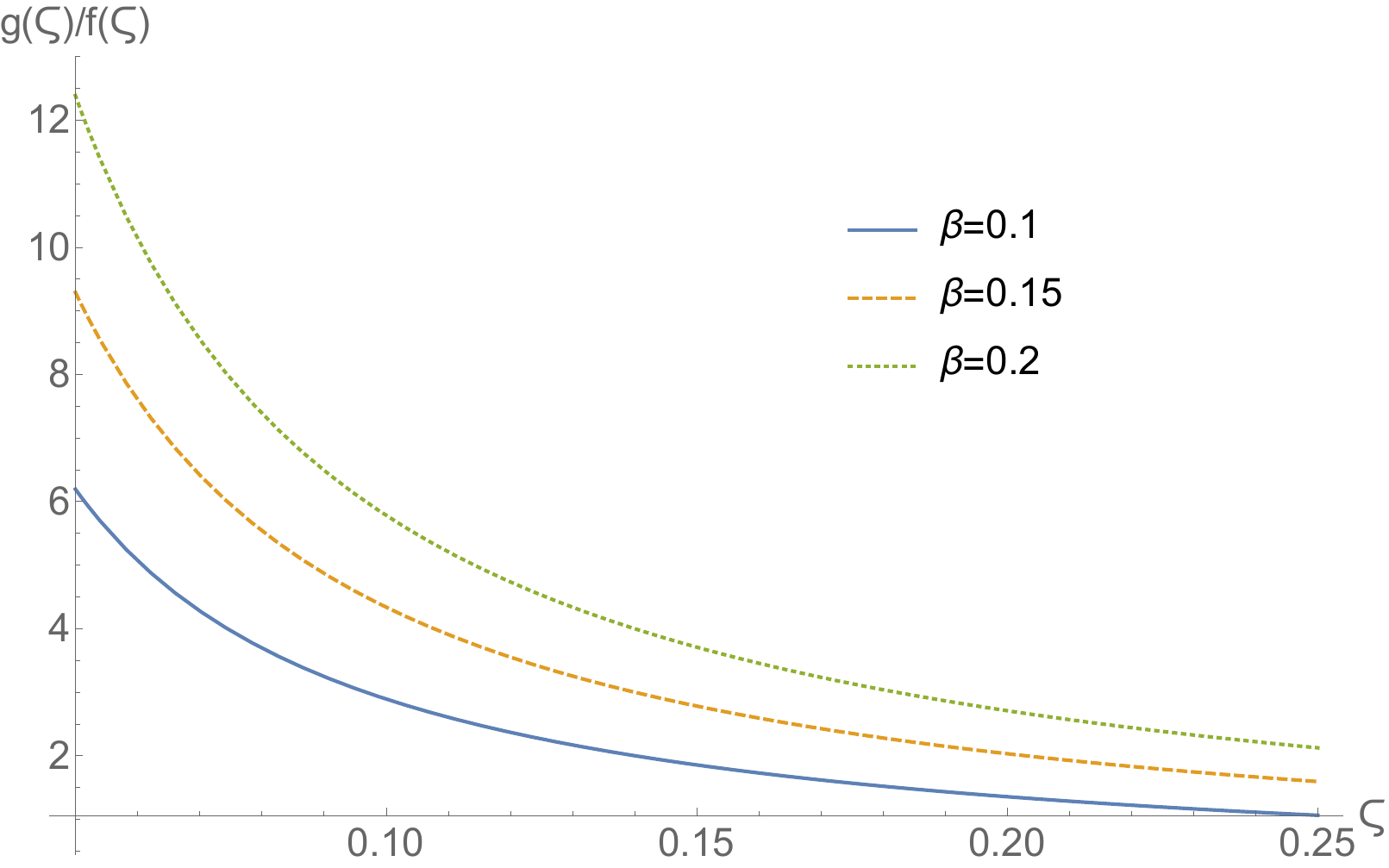}
\caption{Plot of the ratio of dimensionless function $f$ and $g$ from Eq. \eqref{eq:f-g} that define drag resistivity in Eq. \eqref{eq:rhoD} versus $\varsigma$ for several representative values of $\beta$ shown on the plot legends per definition of variables given in Eq. \eqref{eq:variables}.}\label{fig:f-g}
\end{figure}

Finally, by combining Eq. \eqref{eq:D-1-2} with Eqs. \eqref{eq:F} and \eqref{eq:rho}, we obtain the drag resistivity. The remaining frequency and momentum integrals cannot be evaluated in a closed analytical form. However, by rescaling these integrals using the dimensionless units introduced above, we can express the result in a form that lends itself to convenient numerical analysis. These steps lead us to the main result of the paper
\begin{equation}\label{eq:rhoD}
\rho_{\text{D}}=\frac{\sigma}{4\pi^2e^4}\left(\frac{1}{nd^2}\right)^2\frac{T}{E_{\text{F}}}[f(\varsigma)+\chi g(\varsigma)],
\end{equation}  
where $E_{\text{F}}$ is the Fermi energy and an overall factor of temperature $T$ comes from the strength of Langevin fluxes. The dimensionless functions are defined by the following integrals  
\begin{subequations}\label{eq:f-g}
\begin{align}
&f(\varsigma)=\int\limits^{\infty}_{0}\frac{2\varsigma x^5y^4e^{-2x}dxdy}{\pi_+(x,y)\pi_-(x,y)}, \\
&g(\varsigma)=\int\limits^{\infty}_{0}\frac{2\beta\varsigma^2 x^9y^4e^{-2x}dxdy}{\pi_+(x,y)\pi_-(x,y)}\left[\frac{(1+e^{-x})^2}{\pi_+(x,y)}+\frac{(1-e^{-x})^2}{\pi_-(x,y)}\right],
\end{align}
\end{subequations}
where 
$\pi_\pm(x,y)=(y^2-x(1\pm e^{-x}))^2+\varsigma^2y^2x^2(1\pm e^{-x})^2$, 
and dimensionless parameters are given by 
\begin{equation}\label{eq:variables}
\varsigma=\frac{2\pi\sigma}{d\varkappa\omega_p},\quad\beta=\frac{\nu}{d^2\omega_p},\quad\chi=\frac{\alpha}{en}(\bm{j}\cdot\bm{B}). 
\end{equation}
In the more general case, the nonreciprocity parameter $\chi$ entering the drag resistivity can be expressed through different contractions of the nonreciprocity tensor $N_{ijklm}$ over its indices. 
In the limit of vanishing nonreciprocity, $\chi\to0$, Eq. \eqref{eq:rhoD} reduces to the known results for the plasmon-enhanced hydrodynamic drag resistance in conductors laking Galilean invariance \cite{Patel:2017,Zverevich:2023}. 

On Fig. \ref{fig:f-g} we plot the ratio of $g$ and $f$, which quantifies nonreciprocal contribution to drag resistance relative to its main linear part. These functions appear to be numerically close to each other throughout the broad range of relevant parameters. The perturbation theory employed in finding this solutions is controlled by the assumed smallness of the nonreciprocity parameter $\chi<1$. 

In order to justify the choice for the numerical range of parameters we look into their estimates. For this purpose we recall the convention for electron gas parameter $r_s=e^2/v_{\text{F}}\varkappa$. The characteristic plasma frequency is $\omega_p\sim\sqrt{r_s}E_{\text{F}}/\sqrt{k_{\text{F}}d}$, therefore $\varsigma\sim\frac{\sigma}{e^2}\sqrt{r_s/(k_{\text{F}}d)}$. For the interlayer spacing $d\sim200$ nm and the electron density in the range $n\sim 10^{10}\div10^{12}$ cm$^{-2}$ we get $k_{\text{F}}d\sim 2\div20$. For the typical Fermi velocity $v_{\text{F}}\sim10^{6}$ m/s and dielectric constant $\varkappa\sim10$ the electron liquid is weakly correlated $r_s\sim1$, and provided $\sigma/e^2\sim1$ we conclude that $\varsigma\lesssim1$. The estimation of $\beta$ requires further assumptions about the state of electron liquid. For the Fermi liquid regime at $r_s\sim1$ the shear viscosity can be estimated as $\eta\sim n(E_{\text{F}}/T)^2$. The hydrodynamic description of plasmons requires the condition $\omega_p\tau_{\text{ee}}\lesssim1$ which puts a lower bound on the temperature range $E_{\text{F}}/\sqrt[4]{k_{\text{F}}d}<T<E_{\text{F}}$. This means that $\beta\lesssim 1/(k_{\text{F}}d)$. 

In closing, we note that although our analysis is based on perturbation theory, the drag resistance is not necessarily a small perturbative effect. In fact, it can be comparable in magnitude to its linear response value.
For that we need the nonreciprocity parameter $\chi$ to be of the order of unity. In the model considered here $\chi=\alpha Bv$. From the reported experimental result \cite{Pop:2010}, the saturation velocity measurements values of drift velocity in graphene can reach $v\sim 10^5$ m/s, which limits $\alpha B<10^{-5}$ s/m. For particle density at $n\sim 10^{10}$ cm$^{-2}$ the current density $j=env\sim 1$ Cms$^{-1}$ so that for the sample of micron size $L\sim 1\mu$m the total required current would be in a range of micro-Ampere $I\sim 1\mu$A \footnote{The linear response data for transport in graphene typically corresponds to currents in that range $<1\mu$A. Imaging of current flows in graphene devices were reported for total currents in excess of $100\mu$A \cite{Ku:2020}. 
We conclude that nonreciprocal term becomes significant even within the domain of parameters corresponding to the linear response.}. The same estimate also applies to regimes of higher density and lower drift velocity, as long as their product remains approximately constant \footnote{It is perhaps also worth noting that even at the highest drift velocities of the electron fluid, the corresponding Reynolds number, $\mathcal{R}=Lv/\nu$, remains low--specifically, 
$\mathcal{R}\lesssim1$ when using experimentally extracted values for the kinematic viscosity, 
$\nu\sim 0.1$ m$^2$/s \cite{Bandurin:2016}. This justifies the neglect of the nonlinear convective terms in the Navier–Stokes equation.}.

In summary, we have described a possible symmetry-breaking mechanism for the nonreciprocal Coulomb drag effect in electron bilayers and presented supporting calculations within the hydrodynamic regime of the electron liquid. The drag resistance is shown to critically depend on the breaking of Galilean invariance and exhibits a strong temperature dependence through the dissipative properties of the fluid, particularly its intrinsic conductivity and shear viscosity. We provided estimates for Fermi liquids in noncentrosymmetric conductors; however, the generality of the hydrodynamic approach allows for applications to electron systems that do not fall into the paradigm of Fermi liquid theory.

We thank Anton Andreev and Dominique Laroche for many stimulating discussions and prior collaboration on related topics that led to this work. This work was supported by the National Science Foundation Grant No. DMR-2452658 and H. I. Romnes Faculty Fellowship provided by the University of Wisconsin-Madison Office of the Vice Chancellor for Research and Graduate Education with funding from the Wisconsin Alumni Research Foundation.

All data presented in the figures were generated from analytical expressions derived and defined in the paper. The Mathematica code used to produce the plots will be made available by the authors upon reasonable request.

\bibliography{biblio}

\end{document}